\documentstyle[12pt]{article}
\begin{document}
\textheight 22.5cm
\centerline{\large \bf Exact Solutions in Locally Anisotropic} \vskip4pt
\centerline{\large\bf  Gravity and Strings  } \vskip10pt
\centerline{\large \sf Sergiu I. Vacaru} \vskip8pt {\small
\centerline{\noindent{\em Institute of Applied Physics, Academy of Sciences,}}
\centerline{\noindent{\em 5 Academy str., Chi\c sin\v au 2028,
Republic of Moldova}} \vskip5pt
\centerline{\noindent{ Fax: 011-3732-738149, E-mail: vacaru@lises.as.md}}
\centerline{--- --- --- --- ---}
\centerline{\noindent{\em Institute for Basic Research,}}
\centerline{\noindent{\em P. O. Box 1577, Palm Harbor, FL 34682, U. S. A.}}
\vskip3pt \centerline{\noindent{ibr@gte.net,\ http://home1.gte.net/ibr}} }
\vskip20pt

\begin{abstract}
In this Report we outline some basic results on generalized
Finsler--Kaluza--Klein gravity and  locally anisotropic
strings. There are investigated exact solutions for locally
anisotropic Friedmann--Robertson--Walker universes and three dimensional and
string black holes with generic anisotropy.
\end{abstract}

\centerline{\framebox
{ \sf Devoted to the memory of Professor Ryszard Raczka
 (1931--1996)}} \vskip20pt

\section{Introduction}

The theory of locally anisotropic field interactions and (super)strings is
recently descussed \cite{v1,v2,v3,v4} in the context of development of
unified approaches to generalized Finsler like \cite{ma} and Kaluza--Klein
gravity \cite{ow}. A number of present day cosmological models are
constructed as higher--dimensional extensions of general relativity with a
general anisotropic distribution of matter and in correlation with
low--energy limits of string perturbation theory. In non--explicit form it
is assumed the {\bf postulate:\ the matter always (even being anisotropic)
gives rise to a locally isotropic geometry,} which is contained in the
structure of Einstein equations for metric $g_{ij}(x^{k})$ on
(pseudo)Riemannian spaces:
$$
\begin{array}{ccc}
G_{ij}(x^{k}) & \simeq & T_{ij}(x^{k},y^{a}) \\
\mbox{\sf Einstein tensor} &  & \mbox{\sf Energy--momentum tensor }
\\ \fbox{{\sf ( for a locally isotropic curved space)}} &  & \fbox{{\sf (in
general  anisotropic)}}
\end{array}
$$
where $x^i, i=0,1,...,n-1$ are coordinates on space--time $M$ and $y^a,
a=1,2,...,m$ are parameters (coordinates) of anisotropies.

Anisotropic cosmological and locally anisotropic self--gravitating models
are widely used in order to interpret the observable anisotropic structure
of the Universe and of background radiation. Our basic idea to be developed
in this paper is that cosmological anisotropies are not only consequences of
some anisotropic distributions of matter but they reflect a generic
space--time anisotropy induced after reductions from higher to lower
dimensions and by primordial quantum field fluctuations. If usual
Kaluza--Klein theories routinely require compactification mechanisms, we
suggest a more general scenarios of possible decompositions of higher
dimensional (super)space into lower dimensional ones being modelled by a
specific "splitting field" defined geometrically as a nonlinear connection.

A geometry of manifolds provided with a metric more general than the usual
Riemann one, $g_{ij}(x^k)\Longrightarrow g_{ij}(x^k,\lambda ^sy^n),$ where $%
y^n\simeq \frac{dx^n}{dt}$ and $\lambda ^s$ is a parameter of homogeneity of
order $s,$ was proposed in 1854 by B. Riemann and it was studied for the
first time in P. Finsler (1918) and E. Cartan (1934) (see historical
overviews, basic results and references in \cite{ma,v3,v4}). At first sight
there are very substantial objections of physical character to generalized
Finsler like theories:\ One was considered that a local anisotropy crucially
frustrates the local Lorentz invariance. Not having even local
(pseudo)rotations and translations it is an unsurmountable problem to define
conservation laws and values of energy--momentum type, to apply the concept
of fundamental particles fields (for example, without local rotations we can
not define local groups and algebras and their representations). A
difficulty with Finsler like gravity was also the problem of its
inclusion into the framework of modern approaches based on (super)strings,
Kaluza--Klein and gauge theories.

The main purpose of a series of our works (see \cite{v1,v2,v3,v4} and
references) is the development of a general approach to locally
anisotropic gravity imbedding both type of Kaluza--Klein and Finsler--like
theories. It should be emphasized that a subclass of  such models
can be constructed as to have a local space--time Lorentz invariance. We
proved that the general higher order anisotropic gravity can be treated as
alternative low energy limits of (super)string theories with a dynamical
reduction given by the nonlinear connection field and that there are natural
extensions of the Einstein gravity to locally anisotropic theories
constructed on generic nonholonomic vector bundles provided with nonlinear
connection structure.

The field equations of locally anisotropic gravity are of type
$$
\fbox{%
$
G_{\alpha \beta }(x^\alpha ,y^\beta ) \simeq T_{\alpha \beta }(x^\alpha ,
y^\beta ) $}
$$
where the Einstein tensor is defined on a bundle (generalized Finsler)
space, $x^\alpha $ are usual coordinate on the base manifold and $y^\beta $
are coordinates on the fibers (parameters of anisotropy), in general $\dim
\{x^\alpha \}\neq \dim \{y^\beta \}.$

This paper is organized as follows. In Sec. II we briefly review the
geometric background of locally anisotropic gravity. Models of locally
anisotropic Friedmann--Robertson--Walker universe are considered in Sec.
III. In Sec. IV we  analyze anisotropic black hole solutions in three
 dimensional space--times and  extend such
solutions to the string theory. Conclusions are drawn in Sec. V.

\section{Generalized Finsler--Kaluza--Klein gravity}

In Einstein gravity and its locally isotropic modifications of
 Kaluza--Kle\-in,\ 
lower di\-men\-si\-on\-al,
 or of Einstein--Cartan--Weyl types,\ the fun\-da\-men\-tal
spa\-ce--time is considered as a real $(4 +d)$--dimensional,
 where $(d = -2,-1, 0, 1,$ $
..., n),$ manifold of necessary smoothly class and signature, provided with
independent metric (equivalently, tetrad) and linear connection (in general
nonsymmetric). In order to model spaces with generic local anisotropy
instead of manifolds one considers vector, or tangent/cotangent, bundles
(with possible higher order generalizations) enabled with nonlinear
connection and distinguished (by the nonlinear connection) linear connection
and metric structures. The coordinates in fibers are treated as
 parameters of
possible anisotropy and/or as higher dimension coordinates which in general
are not compactified.

In this section we outline the basic results from the so--called locally
anisotropic (la) gravity \cite{ma,v1,v2,v3,v4} (in brief we shall use
la--gravity, la--space and so on).

Let ${\cal E}=(E,\pi ,F,Gr,M)\;$ be a locally trivial vector bundle
(v--bundle) over a base $M$ of dimension $n,$ where $F={\cal R}^m$ is the
typical real
 vector space of dimension $m,$ the structural group is taken to be the
group of linear transforms of ${\cal R}^m,$ i. e. $Gr=GL(m,{\cal R}).$ We
locally parametrize ${\cal E}$ by coordinates $u^\alpha =(x^i,y^a),$ where $%
i,j,k,l,m,...,=0,1,...,n-1$ and $a,b,c,d,...=1,2,...,m.$ Coordinate
transforms $(x^k,y^a)\rightarrow (x^{k^{\prime }},y^{a^{\prime }}) $ on $%
{\cal E},$ considered as a differentiable manifold, are given by formulas $%
x^{k^{\prime }}=x^{k^{\prime }}(x^k),y^{a^{\prime }}= M_a^{a^{\prime
}}(x)y^a,$ where $rank(\frac{\partial x^{k^{\prime }}}{\partial x^k})=n$ and
$M_a^{a^{\prime }}(x)\in Gr.$

One of the fundamental objects in the geometry of la--spaces is the
{\bf nonlinear connection,} in brief
{\bf N--connection.} The N-connection can be
defined as a global decomposition of v-bundle ${\cal E}$ into horizontal, $%
{\cal HE},$ and vertical, ${\cal VE},$ subbundles of the tangent bundle $%
{\cal TE}, {\cal TE}={\cal HE}\oplus {\cal VE}.$ With respect to a
N--connection in ${\cal E}$ one defines a covariant derivation operator $%
\nabla _YA=Y^i\left\{ \frac{\partial A^a}{\partial x^i}+N_i^a(x,A)\right\}
s_a,$ where $s_a$ are local linearly independent sections of ${\cal E},\Lambda
=\Lambda ^as_a$ and $Y=Y^is_i$ is the decomposition of a vector field $Y$
 with respect to a
local basis $s_i$ on $M.$ Differentiable functions $N_i^a(x,y)$ are called
 the coefficients of the N--connection. One holds these transformation
laws for components $N_i^a$ under coordinate transforms:\ $N_{i^{\prime
}}^{a^{\prime }}\frac{\partial x^{i^{\prime }}}{\partial x^i}=M_a^{a^{\prime
}}N_i^a+\frac{\partial M_a^{a^{\prime }}}{\partial x^i}y^a. $ The
N--connection is also characterized by its {\bf curvature}
$$
{\Omega }_{ij}^a=\frac{\partial N_j^a}{\partial x^i}-\frac{\partial N_i^a}{%
\partial x^j}+N_j^b\frac{\partial N_i^a}{\partial y^b}-N_i^b\frac{\partial
N_j^a}{\partial y^b},
$$
and by its linearization which is defined as $\Gamma _{.bi\;}^a(x,y)=\frac{%
\partial N_i^a(x,y)}{\partial y^b}. $ The usual linear connections \ $\omega
_{.b}^a=K_{.bi}^a(x)dx^i $ in a v--bundle ${\cal E}$ form a particular class
of N--connections with coefficients parametrized as $%
N_i^a(x,y)=K_{.bi}^a(x)y^b. $

Having introduced in a v--bundle ${\cal E}$ a N--connection structure we
must modify the operation of partial derivation and introduce a locally
adapted (to the N--connection) basis (frame)
$$
\frac \delta {\delta u^\alpha }=(\frac \delta {\delta x^i}=\partial
_i-N_i^a(x,y)\frac \partial {\partial y^a},\frac \delta {\delta y^a}=\frac
\partial {\partial y^a}),\ \eqno(2.1)
$$
instead of the local coordinate basis $\frac \partial {\partial u^a}=(\frac
\partial {\partial x^i},\frac \partial {\partial y^a}).$ The basis dual to $%
\frac \delta {\delta u^\alpha }$ is written as
$$
\delta u^\alpha =(\delta x^i=dx^i,\delta y^a=dy^a+N_i^a(x,y)dx^i).\eqno(2.2)
$$

We note that a v--bundle provided with a N--connection structure is a
generic nonholonomic manifiold because in general the nonholonomy
coefficients $w_{.\alpha \beta}^{\gamma},$ defined by relations $[\frac
\delta {\delta u^\alpha },\frac \delta {\delta u^\beta }]=\frac \delta
{\delta u^\alpha }\frac \delta {\delta u^\beta }-\frac \delta {\delta
u^\beta }\frac \delta {\delta u^\alpha }=w_{.\alpha \beta }^\gamma \frac
\delta {\delta u^\gamma }, $ do not vanish.

By using bases (2.1) and (2.2) we can introduce the algebra of tensor
distinguished fields (d--fields, d--tensors) on ${\cal E},{\cal C}={\cal C}%
_{qs}^{pr},$ which is equivalent to the tensor algebra of the v--bundle $%
{\cal E}_d$ defined as $\pi _d:{\cal HE}\oplus {\cal VE}\rightarrow {\cal TE}%
 $ \cite{ma}.
 An element $t\in {\cal C}_{qs}^{pr},$ of d--tensor of type $\left(
\begin{array}{cc}
p & r \\
q & s
\end{array}
\right) ,$ are written in local form as%
$$
t=t_{j_1...j_qb_{1...}b_s}^{i_1...i_pa_1...a_r}(u)\frac \delta {\delta
x^{i_1}}\otimes ...\otimes \frac \delta {\delta x^{i_r}}\otimes
dx^{j_1}\otimes ...\otimes dx^{j_p}\otimes 
$$
$$\frac \partial {\partial
y^{a_1}}\otimes ...\otimes \frac \partial {\partial y^{a_r}}\otimes \delta
y^{b_1}\otimes ...\otimes \delta y^{b_s}.
$$

In addition to d--tensors we can consider different types of d-objects with
group and coordinate transforms adapted to a global splitting of v-bundle by
a N--connection.

A {\bf distinguished linear connection,}
 in brief {\bf a d--con\-nec\-ti\-on,} is
defined as a linear connection $D$ in \ ${\cal E}$ conserving as a
parallelism the Whitney sum \ ${\cal HE}\ \oplus \ {\cal VE}$ \ associated
to a fixed N-connection structure in ${\cal E}.$ Components $\Gamma _{.\beta
\gamma }^\alpha $ of a d--connection $D$ are introduced by relations $%
D_\gamma (\frac \delta {\delta u^\beta })=D_{(\frac \delta {\delta u^\gamma
})}(\frac \delta {\delta u^\beta })=\Gamma _{.\beta \gamma }^\alpha (\frac
\delta {\delta u^\alpha }).$

We can compute in a standard manner but with respect to a locally adapted
frame (2.1), the components of {\bf torsion and curvature of a d--connection}
 $D:$  $$
T_{.\beta \gamma }^\alpha =\Gamma _{.\beta \gamma }^\alpha -\Gamma _{.\gamma
\beta }^\alpha +w_{.\beta \gamma }^\alpha \eqno(2.3)%
$$
and
$$
R_{\beta .\gamma \delta }^{.\alpha }=\frac{\delta \Gamma _{.\beta \gamma
}^\alpha }{\delta u^\delta }-\frac{\delta \Gamma _{.\beta \delta }^\alpha }{%
\delta u^\gamma }+\Gamma _{.\beta \gamma }^\varphi \Gamma _{.\varphi \delta
}^\alpha -\Gamma _{.\beta \delta }^\varphi \Gamma _{.\varphi \gamma }^\alpha
+\Gamma _{.\beta \varphi }^\alpha w_{.\gamma \delta }^\varphi . \eqno(2.4)
$$

The global decomposition by a N--connection induces a corresponding
invariant splitting into horizontal $D_X^h=D_{hX}$ (h--derivation ) and
vertical $D_X^v=D_{vX}$ (v--derivation) parts of the operator of covariant
derivation $D,D_X=D_X^h+D_X^v,$ where $hX=X^i\frac \delta {\delta u^i}$ and $%
vX=X^a\frac \partial {\partial y^a}$ are, respectively, the horizontal and
vertical components of the vector field $X=hX+vX\,$ on ${\cal E}.$

Local coefficients $\left( L_{.jk}^i(x,y),L_{.bk}^a(x,y)\right) $ of
covariant h-derivation $D^h$ are introduced as $D_{\left( \frac \delta
{\delta x^k}\right) }^h\left( \frac \delta {\delta x^j}\right)
=L_{.jk}^i\left( x,y\right) \frac \delta {\delta x^i},\quad D_{\left( \frac
\delta {\delta x^k}\right) }^h\left( \frac \partial {\partial y^b}\right)
=L_{.bk}^a(x,y)\frac \partial {\partial y^a} $ and $D_{\left( \frac \delta
{\delta x^k}\right) }^hf=\frac{\delta f}{\delta x^k}=\frac{\partial f}{%
\partial x^k}-N_k^a\left( x,y\right) \frac{\partial f}{\partial y^a}, $
where $f\left( x,y\right) $ is a scalar function on ${\cal E}.$

Local co\-ef\-fi\-ci\-ents $\left( C_{.jk}^i\left( x,y\right)
,C_{.bk}^a\left( x,y\right) \right) $ of v-de\-ri\-va\-ti\-on $D^v$ are
intro\-du\-ced as $D_{\left( \frac \partial {\partial y^c}\right) }^v\left(
\frac \delta {\delta x^j}\right) =C_{.jk}^i\left( x,y\right) \frac \delta
{\delta x^i},\quad D_{\left( \frac \partial {\partial y^c}\right) }^v\left(
\frac \partial {\partial y^b}\right) =C_{.bc}^a\left( x,y\right)$ and\\
 $D_{\left( \frac \partial 
{\partial y^c}\right) }^vf=\frac{\partial f}{\partial y^c}. $

By straightforward calculations we can express respectively the coefficients
of torsion (2.3) and curvature (2.4) \cite{ma} via h- and v-components
parametrized as ${T^{\alpha}}_{\beta\gamma} = \{ T^i_{.jk}, T_{ja}^i,
T_{aj}^i, T_{.ja}^i, T_{.bc}^a \}$ and\\
 $R_{\beta . \gamma\delta}^{. \alpha}
= \{ R_{h.jk}^{.i}, R_{b.jk}^{.a}, P_{j.ka}^{.i}, P_{b.ka}^{.c},
S_{j.bc}^{.i}, S_{b.cd}^{.a} \} .$

The components of the Ricci d--tensor $R_{\alpha \beta }=R_{\alpha .\beta
\tau }^{.\tau } $ with respect to locally adapted frame (2.2) are as follows:%
$R_{ij}=R_{i.jk}^{.k},R_{ia}=-^2P_{ia}=-P_{i.ka}^{.k},
R_{ai}=^1P_{ai}=P_{a.ib}^{.b},R_{ab}=S_{a.bc}^{.c}.$ We point out that
because, in general, $^1P_{ai}\neq ^2P_{ia}$ the Ricci d--tensor is
nonsymmetric.

Now, we shall analyze the compatibility conditions of N- and
 d--con\-nec\-ti\-ons 
and metric structures on the v--bundle ${\cal E}.$ A metric field on ${\cal E%
},\ G\left( u\right) =G_{\alpha \beta }\left( u\right) du^\alpha du^\beta ,$
is associated to a map $G\left( X,Y\right) :{\cal T}_u{\cal E}\times {\cal T}%
_u{\cal E}\rightarrow R, $ pa\-ra\-met\-riz\-ed
 by a non degenerate symmetric $%
(n+m) \times (n+m)$--matrix with components $\widehat{G}_{ij}=G\left( \frac
\partial {\partial x^i},\frac \partial {\partial x^j}\right) ,\widehat{G}%
_{ia}=G\left( \frac \partial {\partial x^i},\frac \partial {\partial
y^a}\right) \ \mbox{ and} \ \widehat{G}_{ab}=G\left( \frac \partial
{\partial y^a},\frac \partial {\partial y^b}\right) . $ One chooses a
concordance between N--connection and G--metric structures by imposing
conditions ${G}\left( \frac \delta {\delta x^i},\frac \partial {\partial
y^a}\right) =0,$ equivalently, $N_i^a\left( x,u\right) =\widehat{G}%
_{ib}\left( x,y\right)$ 
$ \widehat{G}^{ba}\left( x,y\right) ,$ where $\widehat{G%
}^{ba}\left( x,y\right) $ are found to be components of the matrix $\widehat{%
G}^{\alpha \beta }$ which is the inverse to $\widehat{G}_{\alpha \beta }.$
In this case the metric ${G}$ on ${\cal E}$ is defined by two independent
d--tensors, $g_{ij}\left( x,y\right) $ and $h_{ab}\left( x,y\right),$ and
written as
$$
G\left( u\right) =G_{\alpha \beta }\left( u\right) \delta u^\alpha \delta
u^\beta = g_{ij}\left( x,y\right) dx^i\otimes dx^j+h_{ab}\left( x,y\right)
\delta y^a\otimes \delta y^b. \eqno(2.5)%
$$
The d--connection $\Gamma _{.\beta \gamma }^\alpha $ is compatible with the
d--metric structure $G(u)$ on ${\cal E}$ if one holds equalities $D_\alpha
G_{\beta \gamma }=0.$

Having defined the d--metric (2.5) in ${\cal E}$ we can introduce the scalar
curvature of d--connection ${\overleftarrow{R}}=G^{\alpha \beta }R_{\alpha
\beta }=R+S,$ where $R=g^{ij}R_{ij}$ and $S=h^{ab}S_{ab}.$

Now we can write the Einstein equations for la--gravity
$$
{R}_{\alpha \beta } - {\frac{1}{2}} G_{\alpha \beta} {\overleftarrow{R}} + {%
\lambda} G_{\alpha \beta} = {\kappa}_1 {\cal T}_{\alpha \beta} , \eqno(2.6)
$$
where ${\cal T}_{\alpha \beta}$ is the energy--momentum d--tensor on
la--space, ${\kappa}_1$ is the interaction constant and $\lambda$ is the
cosmological constant. We emphasize that in general the d--torsion does not
vanish even for symmetric d--connections (because of
nonholonomy coefficients $w^{\alpha}_{\beta \gamma}).$ So the d--torsion
interactions plays a fundamental role on la--spaces. A gauge like version of
la--gravity with dynamical torsion was proposed in \cite{v5}. We can also
restrict our considerations only with algebraic equations for d--torsion in
the framework of an Einstein--Cartan type model of la--gravity.

Finally, we note that all presented in this section geometric constructions
contain as particular cases those elaborated for generalized Lagrange and
Finsler spaces \cite{ma}, for which a tangent bundle $TM$ is considered
instead of a v-bundle ${\cal E}.$ We also note that the Lagrange (Finsler)
geometry is characterized by a metric of type (2.5) with components
parametrized as $g_{ij}=\frac 12\frac{\partial ^2{\cal L}}{\partial
y^i\partial y^j}$ $\left( g_{ij}=\frac 12\frac{\partial ^2\Lambda ^2}{%
\partial y^i\partial y^j}\right) $ and $h_{ij}=g_{ij},$ where ${\cal L=L}$ $%
(x,y)$ is a Lagrangian ($\left( \Lambda =\Lambda \left( x,y\right) \right) $
is a Finsler metric) on $TM,$ see details in \cite{ma,v1,v2,v3,v4,v5}. The
usual Kaluza--Klein geometry could be obtained for corresponding
parametrizations of N--connection and metric structures on the background
v--bundle.

\section{Anisotropic\ Friedmann-Robert\-son-\-Wal\-k\-er\
 Uni\-ver\-ses}

In this section we shall construct solutions of Einstein equations (2.6)
generalizing the class of Friedmann--Robertson--Walker (in brief FRW)
metrics to the case of $(n=4,m=1)$ dimensional locally anisotropic space. In
order to simplify our considerations we shall consider a prescribed
N-connection structure of type $N_0^1=n(t,\theta ),N_1^1=0,N_2^1=0,N_3^1=0,$
where the local coordinates on the base $M$ are taken as spherical
coordinates for the Robertson-Walker model, $x^0=t,x^1=r,x^2=\theta
,x^3=\varphi ,$ and the anisotropic coordinate is denoted
 $y^1\equiv y.$

The la--metric (2.5) is parametrized by the anzats%
$$
\delta s^2=ds_{RW}^2+h_{11}(t,r,\theta ,\varphi ,y)\delta y^2\eqno(3.1)
$$
where the Robertson-Walker like metric $ds_{RW}^2$ is written as
$$
ds_{RW}^2=-dt^2+a^2(t,\theta )\left[ \frac{dr^2}{1-kr^2}+r^2\left( d\theta
^2+\sin ^2\theta \cdot d\varphi ^2\right) \right] ,
$$
$k=-1,0$ and $1,$ respectively, for open, flat and closed universes, $%
H(t,\theta )=\partial a(t,\theta )/\partial t$ is the anisotropic on angle $%
\theta $ (for our model) Hubble parameter, the containing the $N$-connection
coefficients value $\delta y,$ see (2.2), is of type
$\delta y=dy+n(t,\theta )dt $
and coefficients $n(t,\theta )$ and $h_{11}(t,r,\theta ,\varphi ,y)$ are
considered as arbitrary functions, which are prescribed on la--spaces
defined as  nonholonomic manifolds (in self--consistent dynamical field
 models one must find solutions of a closed system of  equations
for N- and d--connection and d--metric structure).

Considering an anisotropic fluctuation of matter distribution of type \\ 
${\cal %
T}_{\alpha \beta }=T_{\alpha \beta }^{(a)}+T_{\alpha \beta }^{(i)},$ with
nonvanishing anisotropic components $T_{10}^{(a)}(t,r,\theta )\neq 0$ and $%
T_{20}^{(a)}(t,\theta )\neq 0$ and diagonal isotropic energy-momentum tensor
\\ 
$T_{\alpha \beta }^{(i)}=diag(-\rho ,p,p,p,p_{(y)}),$ where $\rho $ is the
matter density, $p$ and $p_{(y)}$ are respectively pressures in 3 dimensional
space and extended space, we obtain from the Einstein equations (2.6) this
generalized system of Friedmann equations:%
$$
\left( \frac 1a\frac{\partial a}{\partial t}\right) ^2=\frac{8\pi G_{(gr)}}%
3\rho -\frac k{a^2},\eqno(3.2)
$$
$$
\frac 1a\frac{\partial ^2a}{\partial t^2}-n(t,\theta )\frac 1a\frac{\partial
a}{\partial t}=-\frac{4\pi G_{(gr)}}3(\rho +3p)\eqno(3.3)
$$
with anisotropic additional relations between nonsymmetric, for la--spaces,
Ricci and energy--momentum d--tensors:%
$$
R_{10}=-n(t,\theta )\left( \frac{kr}{1-kr^2}+\frac 2r\right) \simeq
T_{10}^{(a)}(t,r,\theta ),$$ $$ R_{20}=-n(t,\theta )\cdot ctg\theta \simeq
T_{20}^{(a)}(t,\theta )
$$
when $R_{01}=0$ and $R_{02}=0.$ The $G_{(gr)}$ from (3.2) and (3.3) is the
usual gravitational constant from the Einstein theory.

For the locally isotropic FRW model, when $\rho =-p,$ the
equations (3.2) and (3.3) have an exponential solution of type
$a_{FRW}^{(\exp )}=a_0\cdot e^{\omega _\rho \cdot t}, $
where $a_0=const$ and $\omega _\rho =\sqrt{\frac{8\pi G_{(gr)}}3\rho }.$
This fact is widely applied in modern cosmology.

Substituting (3.2) into (3.3) we obtain the equation
$$
\frac{\partial ^2a}{\partial t^2}-n(t,\theta )\frac{\partial a}{\partial t}%
-\omega _\rho ^2a=0\eqno(3.4)
$$
where the function $a(t,\theta )$ depends on coordinates $t$ and  (as on
a parameter)  $\theta .$ Introducing a new variable $u=a\cdot
\exp \left[ -\frac 12\int n(t,\theta )dt\right] $ we can rewrite the (3.4)
as a parametric equation
$$
\frac{d^2u(t,\theta )}{dt^2}-\widetilde{\omega }(t,\theta )u(t,\theta )=0
$$
for $\widetilde{\omega }(t,\theta )=\omega _\rho ^2+\left( \frac n2\right)
^2+\frac 12\frac{\partial n}{\partial t}$ which admits expressions of the
general solution as series (see \cite{k}).

It is easy to construct exact solutions and understand the physical
properties of the equations of type (3.4) if the  nonlinear connection
structure does not depend on time variable, i.e. $n=n(\theta ).$ By
introducing the new variable $\tau =\omega _\rho t$ and function $a=v\cdot
\exp \left( -D_0\left( \theta \right) \tau \right) ,$ where $D_0\left(
\theta \right) =-n(\theta )/2\omega _\rho ,$ we transform (3.4) into the
equation%
$$
\frac{d^2v}{d\tau ^2}+\left( 1-D_0^2\left( \theta \right) \right) v=0
$$
which can be solved in explicit form:%
$$
v=\left\{
\begin{array}{ccc}
C\cdot e^{-D_0\left( \theta \right) \tau }\cdot \cos (\zeta \tau -\tau _0),
& \quad \varsigma ^2=1-D_0^2\left( \theta \right) , & D_0\left( \theta
\right) <1; \\
C\cdot ch\left( \varepsilon \tau +\tau _0\right) , & \varepsilon
^2=D_0^2\left( \theta \right) -1, & D_0\left( \theta \right) >1; \\
e^{-\tau }\left[ v_0\left( 1+\tau \right) +v_1\tau \right], &  & D_0\left(
\theta \right) \to 1,
\end{array}
\right\}  \eqno(3.5)
$$
where $C, \tau _0, v_0$ and $v_1$ are integration constants.

It is clear from the solutions (3.5)
 that a generic local anisotropy of space--time (possibly induced
from higher dimensions) could play a crucial role in Cosmology. For some
prescribed values of nonlinear connection components we can obtain
exponential anisotropic acceleration, or damping for corresponding
 conditions, of the inflational
scenarios of universes, for another ones there are possible oscillations.

\section{ Anisotropic Black Holes and Strings}

\subsection{Three dimensional la--solutions}

We first consider the simplest possible case when (2+1)--dimensional
space--time admits a prescribed N--connection structure. The anzats for
la--metric (2.5) is chosen in the form%
$$
\delta s^2=-N_{*}^2(r)dt^2+S_{*}^2(r)dr^2+P_{*}^2(r)\delta y^2,\eqno(4.1)
$$
where $\delta y=d\varphi +n(r)dr. $
The metric (4.1) is written for a la--space with local coordinates $%
x^0=t,x^1=r$ and fiber coordinate $y^1=\varphi $ and has components: $%
g_{00}=-N_{*}^2(r),g_{11}=S_{*}^2(r)$ and $h_{11}=P_{*}^2(r).$ The
prescription for N-connection from (2.2) is taken  $N_0^1=0$ and $%
N_1^1=n\left( r\right) .$

The Einstein equations (2.6) are satisfied if one holds the condition
$n=\frac{{\ddot N}_*}{{\dot N}_*}-\frac{{\dot S}_*}{S_{*}},$ 
where, for instance, ${\dot S}_*=\frac{dS_{*}}{dt}.$ So on a
(2+1)--dimensional space--time with prescribed generic N--connection there
are possible nonsingular la--metrics.

Nevertheless (2+1)--like black hole solutions with singular anisotropies
 can be constructed, for instance, by  choosing the parametrizations
$$
P_{*}^2(r)=P^2(r)=\rho ^2\left( r\right) ,\quad N_{*}^2(r)=N^2(r)=\left(
\frac r\rho \right) \cdot \left( \frac{r^2-r_{+}^2}l\right) ,\eqno(4.2)
$$
$$
S_{*}^2=S^2=\left( \frac r{\rho N}\right) ^2,\quad n(r)=N^\varphi \left(
r\right) =-\frac J{2\rho ^2}
$$
where
$$
\rho ^2=r^2+\frac 12\left( Ml^2-r_{+}^2\right) ,\quad r_{+}^2=Ml^2\sqrt{%
1-\left( \frac J{Ml}\right) ^2}
$$
and $J,M,l$ are constants characterizing some values of  rotational
momentum, mass and fundamental length type. In this case the la--metric (4.1)
transforms in the well known BTZ--solution for three dimensional black holes
\cite{btz}.

We can also parametrize solutions for la--gravity of type (4.1) as to be
equivalent to a locally isotropic anti--de Sitter space with cosmological
constant $\Lambda =-\frac 1{l^2}$ when coefficients (4.2) are modified by
 the relations
$N(r)=N^{\perp }=f=\left( -M+\frac{r^2}{l^2}+\frac{J^2}{4r^2}\right)
^{1/2},N^\varphi \left( r\right) =-\frac J{2r^2}, $
where $M>0$ and $\left| J\right| \leq Ml$ and the solution has an outer
event horizon at $r=r_{+}$ and inner horizon at $r=r_{-},$
$r_{\pm }^2=\frac{Ml^2}2\{1\pm \sqrt{1-\left( \frac J{Ml}\right) ^2}\}.
$ We conclude that the N--connection could model both singular
 and nonsingular  anisotropies of (2+1)--dimensional space--times.

\subsection{Three dimensional la--solutions and strings}

We proceed to study the possibility of imbedding of
 3--dimensional solutions of
la--gravity into the low energy dynamics of la--strings \cite{v2,v3,v4}.

A la--metric
$$
\delta s^2=-K^{-1}(r)f(r)dt^2+f^{-1}(r)dr^2+K(r)\delta y^2,\eqno(4.3)
$$
where $\delta y=dx_1+n(r)dt,$ i.e. $N_0^1=n(r)$ and $N_0^1=0,$ solves the
Einstein la--equations (2.6) if
$$
n(r)=\frac 34\varsigma \left( r\right) +\frac{\dot \varsigma (r)}\varsigma
\eqno(4.4)
$$
with $\varsigma (r)=\dot f/f-\dot K/K,$ where, for instance $\dot f=df/dt.$

Metrics of type (4.3) are considered \cite{sf} in an isotropic manner in
connection to solutions of type IIA supergravity that describes a
non--extremal intersection of a solitonic 5--brane, a fundamental string and
a wave along one of common directions.

We have an anisotropic plane wave solution in $D+1$ dimensions if%
$$
\delta s^2=-K^{-1}(r)f(r)dt^2+f^{-1}(r)dr^2+r^2d\Omega _{D-2}^2+K(r)\delta
y_{(\alpha )}^2\eqno(4.5)
$$
where $
\delta y_{(\alpha )}=dx_1+\left[ 1/K^{\prime }\left( r\right) -1+\tan \alpha
\right] dt, $\
$K(r)=1+\mu ^{D-3}\sinh ^2\alpha /r^{D-3},$ $\left( K^{\prime }(r)\right)
^{-1}=1-\mu ^{D-3}\sinh \alpha \cosh \alpha /(r^{D-3}K),f\left( r\right)
=1-\mu ^{D-3}/r^{D-3}$ for isotropic solutions but $f(r)$ is a function
defined by the prescribed component of N--connection (4.4) for la--spaces, $%
r^2=x_2^2+...x_D^2,$ and the parameter $\alpha $ define shift translations.

A la--string \cite{v2,v3,v4} solution is constructed by including (4.5) into
a 10--dimensional la--metric with trivial shift $\delta y_{(\alpha =0)}$
$$
\delta s_{(10)}^2=H_f^{-1}\left[ -\frac{f(r)}{K(r)}dt^2+K(r)\delta
y_{(0)}^2\right] + $$
$$ dx_2^2+...+dx_5^2+H_{S5}\left[ f^{-1}(r)dr^2+r^2d\Omega
_3^2\right]
$$
where the la--string dilaton fields and antisymmetric tensor are
 defined \cite{sf}
by the relations $e^{-2\phi }=H_{S5}^{-1}H_f,B_{01}=H_f^{-1}+\tanh \alpha _f,
$ $r^2=x_6^2+...x_8^2,$ $H_{ijk}=\frac 12\epsilon _{ijkl}\delta _lH_{S5}$
(in general, one considers la--derivations of type (2.2)) and $%
i,j,d,l=6,...,9.$

Considering  dimensional reductions in variables $x_1,x_2,x_3,x_4,x_5$
one can construct  non--extremal, under singular anisotropies, 5--dimensional
la--black hole solutions
$$
\delta s_{(5)}^2=-\lambda ^{-2/3}f(r)dt^2+\lambda ^{1/3}\left[
f^{-1}(r)dr^2+r^2d\Omega _3^2\right]
$$
where $\lambda =H_{S5}H_fK=\left( 1+\frac{Q_{S5}}{r^2}\right) \left( 1+\frac{%
Q_f}{r^2}\right) \left( 1+\frac{Q_K}{r^2}\right) ;Q_{S5},Q_f$ and $Q_K$ are
constants. Finally we note that for la--backgrounds the
 function  $f\left( r\right) $ is
connected with the components of N--connection via relation (4.4), i.e.
 the N--connection structure could model both type of singular
 (like black hole ) and nonsingular locally anisotropic string solutions.

\section{Conclusions}
 The scenario of modelling  of physical theories with generic
 locally anisotropic interactions on nonholonomic
 bundles provided with nonlinear connection structure
 has taught us a number of interesting things about
 a new class of anisotropic cosmological models,  black hole solutions
 and  low energy limits of string theories. Generic anisotropy
 of space--time could be a consequence of  reduction from
 higher  to lower dimensions  and of quantum filed
 and space--time structure fluctuations in pre--inflationary
 period. This way an unification of logical aspects, geometrical
 background  and physical ideas from the generalized Finsler
 and Kaluza--Klein theories was achieved.

The focus of this paper was to present some exact solutions
 with prescribed nonlinear connection
 for the locally anisotropic gravity and string theory.
 We have shown that a generic anisotropy of
 Friedmann--Robertson--Walker metrics could result in
 drastic  modifications of cosmological models. It was our task
 here to point the conditions when the nonlinear connection
 will model singular, or nonsingular, anisotropies with
 three dimensional black hole solutions and to investigate
 the possibility of generalization of such type constructions
 to string theories.

\vskip15pt
{\bf Acknowledgments}
\vskip5pt
The author would like to thank the Organizing Committee of International
Conference "Particle, Fields and Gravitation", Lodz, April 15--19, 1998, for
support of his participation. He is very grateful to Profs. A. Trautman and
S. Bazanski for hospitality during his visit to Warsaw University.

\end{document}